\documentstyle[pramana,epsf,floats]{ias}

\newcommand{\To}{\longrightarrow}

\newcommand{\be}{\begin{equation}}
\newcommand{\ee}{\end{equation}}
\newcommand{\bea}{\begin{eqnarray}}
\newcommand{\eea}{\end{eqnarray}}
\newcommand{\nn}{\nonumber}

\newcommand{\gud}{g_{\mu\nu }}
\newcommand{\upm}{U^{\mu }}
\newcommand{\udm}{U_{\mu}}
\newcommand{\vpm}{V^{\mu }}
\newcommand{\vdm}{V_{\mu}}
\newcommand{\bab}{B_{\alpha \beta}}
\newcommand{\bagd}{B_{\alpha\gamma}}
\newcommand{\bagu}{B^{\gamma\alpha}}
\newcommand{\sagd}{\Sigma_{\alpha\gamma}}
\newcommand{\sagu}{\Sigma^{\gamma\alpha}}
\newcommand{\oagd}{\Omega_{\alpha\gamma}}
\newcommand{\oagu}{\Omega^{\gamma\alpha}}
\newcommand{\upa}{U^{\alpha }}
\newcommand{\uda}{U_{\alpha}}
\newcommand{\vpa}{V^{\alpha }}

\newcommand{\p}{\partial}

\newcommand{\oab}{\Omega_{\alpha\beta}}
\newcommand{\sab}{\Sigma_{\alpha\beta}}
\newcommand{\fmn}{F_{\mu\nu}}
\newcommand{\hab}{h_{\alpha\beta}}
\newcommand{\uab}{U_{\alpha ; \beta}}
\newcommand{\uba}{U_{\beta ; \alpha}}
\newcommand{\vab}{V_{\alpha ; \beta}}

\newcommand{\upg}{U^{\gamma }}

\newcommand{\rmn}{R_{\mu\nu}}
\newcommand{\tmn}{t_{\mu\nu}}
\newcommand{\gmn}{g_{\mu\nu}}
\title{On a Raychaudhuri equation for hot gravitating fluids}
\author {Chandrasekher Mukku}
\address {International Institute of Information Technology, Hyderabad, Andhra Pradesh, 500 032, India}
\author {Swadesh M. Mahajan}
\address {Institute of Fusion Studies, University of Texas, Austin, Texas,78712 U.S.A}
\author {Bindu A. Bambah}
\address {School of Physics, University of Hyderabad, Hyderabad, Andhra Pradesh, 500 046, India.}

\begin{document}




\abstract {We generalize the  Raychaudhuri equation for the
evolution of a self gravitating fluid to include an Abelian and
non-Abelian hybrid magneto fluid at a finite temperature.  The aim
is to utilize this equation for investigating the dynamics of
astrophysical high temperature Abelian and non-Abelian plasmas.}

\maketitle

\section{Introduction}It is perhaps one of the finest tributes to the simplicity and
elegance of Einstein's theory of Gravitation and to the human spirit,
that a college teacher sitting in Kolkata  was able to conceive the
equation for the evolution of a gravitating fluid now known as
Raychaudhuri's equation \cite{rc}. The equation served as a lemma for the
Penrose-Hawking singularity theorems and for the study of exact
solutions in general relativity \cite{hawking,penrose}. It provides a simple validation of
our expectation that gravitation should be a universal attractive
force between any two particles in general relativity \cite{dadhich}. This equation
has stood the test of time and has been generalized in many ways. It
has found applications in modern theories of strings and membranes \cite{kar,elize}.
 We attempt a more modest
generalization to tackle the statistical properties of a hot
astrophysical plasma. These may be electromagnetic or chromomagnetic
(quark gluon) plasmas. We are guided here by a recent formalism that  has been used
to investigate the dynamics of a hot charged fluid in terms of a
hybrid magneto-fluid and the changes brought about to the
Raychaudhuri equation by the introduction of a statistical attributes associated with
finite temperature \cite{mahajan} are many and interesting. We give also give a brief
outline of how to generalize this to  the evolution of gravitating
non-Abelian plasmas in the early universe \cite{bambah}.

\section{Raychaudhuri's equation for charged gravitating
fluids. }

In this section, for the sake of completeness, we first review the
derivation of the standard Raychaudhuri equation for the evolution
of a self gravitating fluid. We follow the derivation from
Raychaudhuri's original work for geodesic flow \cite{book}. We then
incorporate the Lorentz force  into the Raychaudhuri
equation for non geodesic flow in the presence of an electromagnetic
field.

Consider a self gravitating fluid in which the fluid particles have
a velocity $U_{\mu}$. In a spacetime with metric $g_{\mu\nu}$, these
velocity vectors allow us to define an orthogonal projection vector
\be h_{\mu\nu}=g_{\mu\nu}+U_{\mu}U_{\nu}. \ee The following
properties hold for $h_{\mu\nu}$: \begin{itemize}
\item{1.} $h_{\mu}^{\mu}=g^{\mu}_{\mu}+U_{\mu}U^{\mu}=3$, assuming
$U_{\mu}U^{\mu}=-1$. This implies that $h_{\mu\nu}$ is three
dimensional.
\item{2.}\be \upm h_{\mu\nu}=\upm\gud +\upm\udm U_{\nu}=0 \ee and \be
h_{\mu\nu}U^{\nu}=\gud U^{\nu}+\udm U^{\nu}U^{\nu}=\udm-\udm=0\ee
This implies $\upm$ is orthogonal to $h_{\mu\nu}$. Since $\upm$ is
tangent to timelike curves, $h_{\mu\nu}$ is purely spacelike. It is
the metric of the three dimensional hypersurface orthogonal to
$\upm$.
\end{itemize}

For any time-like vector $\upm$ in a spacetime, the geodesic
deviation equation can be written as \be
\upm{}_{;\alpha\beta}-\upm{}_{;\beta\alpha}=R^{\mu}{}_{\sigma \alpha
\beta}U^{\sigma} \label{gd}\ee

Although Eqn.\ref{gd} is the geodesic deviation equation, it holds
for any timelike vector $\upm$ and not just the tangent vector to
the geodesic. This is an important observation since Electromagnetic
and Yang-Mills fields cause the fluid motion to be non-geodesic. Let
us consider the parallel transport of $\upm$,
$\bab=U_{\alpha;\beta}$. It has the following properties:
\begin{itemize}
\item{a.)}$\upa\bab=\upa
U_{\alpha;\beta}=\frac{1}{2}(\upa\uda)_{;\beta}=0.$
\item{b.)}$\bab U^{\beta}=U_{\alpha;\beta}U^{\beta} \ne 0 $ unless
geodesic.
\end{itemize}
Thus, $\bab$ 
is a purely spatial tensor for
geodesic motion, but fails to be fully spatial for non geodesic
motion. In particular, with an EM field present \be
U_{\alpha;\beta}U^{\beta}=\frac{q}{m}F_{\alpha \beta}U^{\beta} \ee.

Following Raychaudhuri, let us decompose $\bab$ into its irreducible
parts:
\begin{itemize}
\item{i.} The antisymmetric part ; $\oab$,
\item{ii.} The symmetric, trace free part ; $\sab$,
\item{iii.} The trace ; $\frac{1}{3}\Theta \hab$.
\end{itemize}

We can write \be
\bab=-U_{\alpha;\gamma}U^{\gamma}U_{\beta}+\sab+\oab+\frac{1}{3}\Theta\hab
, \ee
with  $\Theta$, giving the rate of dilation of a three space
element locally orthogonal to $\upa$, emerging as the trace
 \be
B^{\alpha}{}_{\alpha}=U^{\alpha}{}_{;\alpha}=\Theta .\ee
We can define \be
\oab=\frac{1}{2}(\uab-\uba)+\frac{1}{2}(\dot{\uda}U^{\beta}-\dot{U_{\beta}}\uda),
\ee where, $\dot{\uda}=\uab U^{\beta}$ is the acceleration vector,
which gives the departure of the velocity field from geodesity.
Similarly
\be\sab=\frac{1}{2}(\uab+\uba)+\frac{1}{2}(\dot{\uda}U^{\beta}+\dot{U_{\beta}}\uda)-\frac{1}{3}\Theta\hab.
\label{sab} \ee Consider
\bea\upa\oab&=&\frac{1}{2}(\upa\uab-\upa\uba)+\frac{1}{2}(\upa\dot{\uda}U^{\beta}-\upa\dot{U_{\beta}}\uda)
\nn \\
&=&-\frac{1}{2}(\uab\upa +\frac{1}{2}(\uba)+\frac{1}{2}(\upa
U_{\alpha;\gamma}U^{\gamma})U_{\beta}+\frac{1}{2}\uba\upa=0, \eea
and \be \oab U^{\beta}=\frac{1}{2}(\uab U^{\beta}-\uba
U^{\beta})+\frac{1}{2}(\dot{\uda}U^{\beta}U_{\beta}-\dot{U_{\beta}}\uda
U^{\beta})=0. \ee These relations imply that $\oab$ is spatially
antisymmetric. Similarly, \bea
\upa\sab&=&\frac{1}{2}(\upa\uab+\upa\uba)+\frac{1}{2}(\upa\dot{\uda}U^{\beta}+\upa\dot{U_{\beta}}\uda)-\frac{1}{3}\Theta\upa\hab
\nn \\ &=&\frac{1}{2}(\dot{U^{\beta}}-\dot{U_{\beta}})=0 .\eea and,
\bea \sab U^{\beta}&=&\frac{1}{2}(\uab U^{\beta}+\uba
U^{\beta})+\frac{1}{2}(\dot{\uda}U_{\beta}U^{\beta}+\dot{U_{\beta}}U^{\beta}\uda)-\frac{1}{3}\Theta\hab
U^{\beta} \nn \\ &=&\frac{1}{2}(\dot{U^{\alpha}}-\dot{U_{\alpha}})=0
.\eea So we have taken care of the non geodesic nature of $\upa$ to
define $\sab$ and $\oab$, by including the acceleration due to the
EM force and made them spatial.


Let us now consider a parallely transported $\bab$, $B_{\alpha
\beta;\gamma}=U_{\alpha;\beta\gamma}$. Eq. \ref{gd} allows us to
write \be B_{\alpha
\beta;\gamma}=(U_{\alpha;\gamma\beta}-R_{\alpha\mu\gamma\beta}U^{\mu})
\ee Thus, \bea B_{\alpha
\beta;\gamma}\upg&=&(U_{\alpha;\gamma\beta}\upg-R_{\alpha\mu\gamma\beta}U^{\mu}\upg)\nn \\
&=&(U_{\alpha;\gamma}\upg)_{;\beta}-U_{\alpha;\gamma}\upg{}_{;\beta}-R_{\alpha\mu\gamma\beta}U^{\mu}\upg \nn \\
&=&(\dot{\uda}){}_{;\beta}-B_{\alpha\gamma}B^{\gamma}{}_{\beta}-R_{\alpha\mu\gamma\beta}U^{\mu}\upg.\eea
Taking the trace over $\alpha$ and $\beta$  we get \be
B^{\alpha}{}_{\alpha ;
\gamma}\upg=(\dot{\upa})_{;\alpha}-B_{\alpha\gamma}B^{\gamma\alpha}-R^{\alpha}{}_{\mu\gamma\alpha}U^{\mu}U^{\gamma}.
\ee But \be B^{\alpha}{}_{\alpha}=\Theta\ee since $\sab$ is
traceless and $\oab$ is antisymmetric. Therefore Eq.16 becomes
\be
\Theta_{;\gamma}\upg=(\dot{\upa})_{;\alpha}-B_{\alpha\gamma}B^{\gamma\alpha}-R_{\mu\gamma}U^{\mu}\upg
\ee where $R_{\mu\gamma}$ is the Ricci tensor. Using the Lorentz
force law equation $ \dot{\upa}=\frac{q}{m}F^{\alpha}{}_{ \gamma}U^{\gamma}$, we have \bea
\dot{\Theta}&=&(\frac{q}{m}F^{\alpha}{}_{\gamma}\upg){}_{;,\alpha}-\bagd\bagu-R_{\mu\nu}\upm U^{\nu} \nn \\
&=&\frac{q}{m}F^{\alpha}{}_{\gamma;\alpha}\upg+\frac{q}{m}F^{\alpha}{}_{\gamma}B^{\gamma}{}_{\alpha}-\bagd\bagu-R_{\mu\nu}\upm U^{\nu}\nn\\
&=&\frac{q}{m}F^{\alpha}{}_{\gamma;\alpha}\upg+\frac{q}{m}F^{\alpha}{}_{\gamma}\Omega
^{\gamma}{}_{\alpha}-\bagd\bagu-R_{\mu\nu}\upm U^{\nu} \eea Where we
have the antisymmetric nature of $F_{\alpha\beta}$, so that
$F_{\alpha\beta}B^{\beta\alpha}=F_{\alpha\beta}\Omega^{\alpha\beta}$.
We also have \be
\bagd\bagu=\sagd\sagu+\oagd\oagu+\frac{1}{9}\Theta^{2}
h_{\alpha\gamma}h^{\gamma\alpha} \ee

Thus \be
\dot{\Theta}=\frac{q}{m}F^{\alpha}{}_{\gamma;\alpha}\upg+\frac{q}{m}F^{\alpha}_{\gamma}\Omega
^{\gamma}_{\alpha}-\sagd\sagu-\oagd\oagu-\frac{1}{3}\Theta^{2}
-R_{\mu\nu}\upm U^{\nu}. \ee
The Ricci  tensor is related to the energy momentum tensor by the  Einstein's equation \be
\rmn-\frac{1}{2}R\gmn=8\pi\tmn \ee or, equivalently, by \be
\rmn=8\pi(\tmn-\frac{1}{2}t \gmn) \ee
 Substituting  for $R_{\mu\nu}$ t we find
  \bea
\dot{\Theta}&=&\frac{q}{m}F^{\alpha}{}_{\gamma;\alpha}\upg+\frac{q}{m}F^{\alpha}{}_{\gamma}\Omega
^{\gamma}{}_{\alpha}-\sagd\sagu-\oagd\oagu\nn\\
&-&\frac{1}{3}\Theta^{2} -8\pi(\tmn-\frac{1}{2}t \gmn)\upm
U^{\nu}.\eea With our notation, ($U_{\mu}U^{\mu}=-1$), the limit
$\fmn\longrightarrow0$ is the standard (geodesic) Raychaudhuri
equation. $\sab$ is called the shear tensor, $\oab$ is called the
vorticity tensor and $\theta$ is called the expansion scalar. These
names arise because one can define an orthogonal triad of vectors in
the three dimensional hypersurface orthogonal to $\upm$. Under
Fermi-Walker transport, the volume defined by the triad expands,
gets deformed and also rotated.

\section{Raychaudhuri's equation for a "unified" charged gravitating fluid}

For the use of the Raychaudhuri equation to describe astrophysical
plasmas in which gravitational effects are compatible with the fluid attributes, we have to account for the temperature and pressure of the
fluid. For this, we incorporate into a small volume element of the
fluid a statistical factor $f$ which represents a temperature
dependent statistical attribute of the fluid, and is related to the
enthalpy $h$, the scalar density in the rest frame $n$ and  the mass $m$ of the
fluid particles  by the relation $ h=mnf(T)$. When one does the kinetic theory of high temperature plasmas $f(T)$ seems to emerge as the most useful variable to represent temperature effects. For relativistic plasmas
 $h=mn\frac{K_{3}(m_s/T)}{K_{2}(m_s/T)} $ and $f(T)$ is purely a function of temperature.The velocity vector of the
fluid is obtained as the average velocity of this small volume of
the fluid and is written as $\vpm=f\upm$. This drastically alters
the character of the terms in the evolution equation. For example,
now $\vpm\vdm=-f^2$ in contrast with $\upm\udm=-1$ and unlike
$\upa\uab=\frac{1}{2}(\upa\uda)_{;\beta}=0, $, now
$\vpa\vab=-f\partial_{\beta }f$. These terms significantly change the
spatial terms of the Raychaudhuri equation and necessitating a generalization
to account for these statistical factors. Indeed, this
statistical  limit  for the fluid velocity \cite{mahajan}, unlike the particle
limit, provides a natural factor for producing acceleration forces
from within the fluid due to pressure and temperature gradients.
In the unified magnetofluid picture one can write the equation of motion of a  magneto-fluid with entropy $\sigma$ as
\be
T\p^{\nu}\sigma=gM^{\mu\nu}U_{\mu}\label{eom3}
\ee
where
\be
M^{\mu \nu}=F^{\mu
\nu}+\frac{m}{g}S^{\mu \nu} \label{eom},
\ee
and
\be
S_{\mu\nu}=\partial_{\mu}(fU_{\nu})-\partial_{\nu}(fU_{\mu}) ,\label{smn}\ee
are antisymmetric second rank "flow" tensors defined in \cite{mahajan}.
In this sense, $M_{\mu\nu}$ represents an anti-symmetric "unified" field -flow tensor constructed from the kinematic ($\upm$) , statistical ($f(T)$) and electromagnetic  ($F_{\mu\nu}$) attributes of the magneto-fluid.

While the statistical factors  provide us with additional
acceleration terms which alter the purely spatial character of the
shear and vorticity tensors, there are some relations  that remain
unaltered. A trivial example is that $\vpm$ remains orthogonal to the hypersurface with
metric $\hab$ defined as the projector $\hab=g_{\alpha\beta}+\uda
U_{\beta}$. To derive the geodesic deviation for $\vpm$ notice that
\bea
\vpm{}_{;\alpha\beta}&=&(f\upm{}_{;\alpha}+\upm f_{;\alpha}){}_{;\beta} \nn \\
&=&f_{;\beta}\upm{}_{;\alpha}+\upm{}_{;\beta}f_{;\alpha}+\upm
f_{;\alpha\beta}+f\upm{}_{;\alpha\beta} .\eea It is easy to see that
anti symmetrization in indices $\alpha$,$\beta$ reduces the geodesic
deviation to \bea
\vpm{}_{;\alpha\beta}-\vpm{}_{;\beta\alpha}&=& f(\upm{}_{;\alpha\beta}+\upm{}_{;\beta \alpha}) \nn\\
&=&fR^{\mu}{}_{\sigma\alpha\beta}U^{\sigma} \nn\\
&=&R^{\mu}{}_{\sigma\alpha\beta}V^{\sigma}, \label{dev}\eea and its
character is unchanged.

The question that arises now is if it is possible to define the
generalizations of the standard definitions of the shear and vorticity tensors in a similar
manner, and can they be constructed to be
purely spatial tensors.

Let us, following the standard procedure, decompose $\tilde{B}_{\alpha\beta}$
into its irreducible parts and define: \be
\tilde{B}_{\alpha\beta}=\tilde{\Sigma}_{\alpha\beta}+\tilde{\Omega}_{\alpha\beta}+\frac{1}{3}\tilde{\Theta}h_{\alpha\beta}+\frac{1}{f}V_{\alpha}\partial_{\beta}f.\ee
Let us examine the trace of $\tilde{B}_{\alpha\beta}$ as this
defines the expansion scalar $\tilde{\Theta}$\be
\tilde{B}^{\alpha}{}_{\alpha}=\tilde{\Theta}+\frac{1}{f}V^{\alpha}\partial_{\alpha}f=V^{\alpha}{}_{;\alpha}\label{theta}\ee
where we have assumed that the generalized shear tensor
$\tilde{\Sigma}_{\mu\nu}$ is symmetric and traceless. If we now
define \be
\tilde{\Sigma}_{\alpha\beta}=\frac{1}{2}(V_{\alpha;\beta}+V_{\beta;\alpha})-\frac{1}{3}\tilde{\Theta}h_{\alpha\beta}-\frac{1}{2f}(V_{\alpha}\partial_{\beta}f+V_{\beta}\partial_{\alpha}f)\label{shear},\ee
then its trace \be
\tilde{\Sigma}^{\mu}{}_{\mu}=V^{\mu}{}_{;\mu}-\tilde{\Theta}-\frac{1}{f}V^{\mu}\partial_{\mu}f\ee
because of equation(\ref{theta}), goes to zero; the trace free condition also reproduces the required definition of $\tilde{\Theta}$.

The generalized vorticity tensor $\tilde{\Omega}_{\alpha\beta}$ may also be written as
\be
\tilde{\Omega}_{\alpha\beta}=\frac{1}{2}(V_{\alpha}{}_{;\beta}-V_{\beta}{}_{;\alpha})-\frac{1}{2f}(V_{\alpha}\partial_{\beta}f-V_{\beta}\partial_{\alpha}f).\ee
We see that the tensor $S_{\mu\nu}$ defined in Eqn.\ref{smn}
allows us the following identification:\be
\tilde{\Omega}_{\alpha\beta}=\frac{1}{2}S_{\beta\alpha}-\frac{1}{2f}(V_{\alpha}\partial_{\beta}f-V_{\beta}\partial_{\alpha}f).\ee

Then, a straight forward analysis shows that \be
V^{\alpha}(\tilde{\Sigma}_{\alpha\beta}+\tilde{\Omega}_{\alpha\beta})=0.\ee
An easy consequence of this result is that \be
V^{\alpha}\tilde{B}_{\alpha\beta}=0.\ee From earlier work, we know
that the gradient of $f$, for a perfect fluid, is related to the
pressure gradient and to Lorentz forces from the presence of a
electromagnetic field \cite{mahajan}. So, unlike the standard particle picture
of the shear and vorticity, the unified picture of shear and
vorticity automatically includes accelerations coming from the
internal forces of the fluids. This in turn ensures that the shear
is now not a purely spatial tensor; the vorticity
tensor behaves in a similar manner. Indeed, we see from the above equation that it is the sum of
the shear and vorticity tensors that allows simplification.

Let us examine $\tilde{B}_{\alpha\beta}V^{\beta}$: Again, it is easy
to see that \be
(\tilde{\Sigma}_{\alpha\beta}+\tilde{\Omega}_{\alpha\beta}+\frac{1}{f}V_{\alpha}\partial_{\beta}f)V^{\beta}=V_{\alpha;\beta}V^{\beta}=\dot{V_{\alpha}},\ee
implying
$\tilde{B}_{\alpha\beta}$ \be V^{\alpha}\tilde{B}_{\alpha\beta}=0\ee
and \be
\tilde{B}_{\alpha\beta}V^{\beta}=V_{\alpha;\beta}V^{\beta}=\dot{V_{\alpha}}.\ee
The decomposition of $V_{\alpha;\beta}$ into its irreducible
components can therefore be written as: \be
V_{\alpha;\beta}=\tilde{B}_{\alpha\beta}=\tilde{\Sigma}_{\alpha\beta}+\tilde{\Omega}_{\alpha\beta}+\frac{1}{3}\tilde{\Theta}h_{\alpha\beta}+\frac{1}{f}V_{\alpha}\partial_{\beta}f.\ee
So, while the generalized expansion scalar, shear and vorticity
tensors show evidence of internal fluid forces, the tensor,
$\tilde{B}_{\alpha\beta}=V_{\alpha;\beta}$ itself shows similarity
with the original tensor, $B_{\alpha\beta}=U_{\alpha;\beta}$ defined
earlier in the particle picture of the gravitating fluid. I.e; \be
V^{\alpha}\tilde{B}_{\alpha\beta}=0,\ee and \be
\tilde{B}_{\alpha\beta}V^{\beta}=V_{\alpha;\beta}V^{\beta}=\dot{V_{\alpha}}.\ee
A careful analysis shows why this is so. From the definitions of
$\tilde{\Theta}$, $\tilde{\Sigma}_{\mu\nu}$ and
$\tilde{\Omega}_{\mu\nu}$, it is easy to see that the following
equations relate the scalar of expansion, shear and vorticity in the
particle view to those in the unified view of gravitating fluids:
\bea \tilde{\Theta}&=&f \Theta\nn\\ \tilde{\Sigma}_{\mu\nu}&=&f
\Sigma_{\mu\nu}\nn \\ \tilde{\Omega}_{\mu\nu}&=&f
\Omega_{\mu\nu},\eea which then imply, \be
\tilde{B}_{\mu\nu}=fB_{\mu\nu}+V_{\mu}\partial_{\nu}lnf .\ee In
analogy with the derivation of the standard Raychaudhuri equation,
if we define $\tilde{B}_{\alpha\beta}=\vab$, then from the deviation
equation (\ref{dev}),\bea
\tilde{B}_{\alpha\beta}{}_{;\gamma}V^{\gamma}&=&(V_{\alpha;\gamma\beta}V^{\gamma}-R_{\alpha\sigma\gamma\beta}V^{\sigma}V^{\gamma} )\nn \\
&=& ((V_{\alpha;\gamma}V^{\gamma}){}_{;\beta}-V_{\alpha;\gamma}V^{\gamma}{}_{;\beta}-R_{\alpha\sigma\gamma\beta}V^{\sigma}V^{\gamma} ) \nn \\
&=&(\dot{V_{\alpha}}){}_{;\beta}-\tilde{B}_{\alpha\gamma}\tilde{B}^{\gamma}{}_{\beta}-R_{\alpha\sigma\gamma\beta}V^{\sigma}V^{\gamma})
\eea Taking the trace over the indices $\alpha \beta$ and using
$\tilde{B}^{\alpha}{}_{\alpha}=V^{\alpha}{}_{;\alpha}=\tilde{\Theta}+\frac{1}{f}V^{\alpha}\partial_{\alpha}f$,
the new expansion scalar, we have \be
(\tilde{\Theta}+\frac{1}{f}V^{\mu}\partial_{\mu}f)_{;\gamma}V^{\gamma}=(\dot{V^{\alpha}}_{;\alpha}-\tilde{B}_{\alpha\gamma}\tilde{B}^{\gamma\alpha}-R_{\sigma\gamma}V^{\sigma}V^{\gamma}).
\ee As before, we use Einstein's equation to get \be
(\tilde{\Theta}+\frac{1}{f}V^{\mu}\partial_{\mu}f)_{;\gamma}V^{\gamma}=(\dot{V^{\alpha}}_{;\alpha}-\tilde{B}_{\alpha\beta}\tilde{B}^{\beta\alpha}-8
\pi (t_{\mu \nu}-\frac{1}{2}t g_{\mu\nu})V^{\mu}V^{\nu}).
\label{re1}\ee

The acceleration forces are twofold, one coming from the em forces
within the fluid and the second from the fluid forces themselves
(pressures,etc). To examine what these are, we return to the
equation of motion  for a fluid with entropy $\sigma$ in the unified picture Eqn.\ref{eom}  and use the fact that \be
S_{\mu\nu}=\partial_{\mu}(f(T)U_{\nu})-\partial_{\nu}(f(T)U_{\nu})=V_{\nu;\mu}-V_{\mu;\nu}\ee
For an isentropic fluid, \be
U^{\mu}\partial_{\mu}\sigma=\sigma_{;\mu}V^{\mu}=0\ee and therefore,
we can write \bea
T\partial_{\nu}\sigma&=&-TU^{\mu}[U_{\mu}\partial_{\nu}\sigma-U_{\nu}\partial_{\mu}\sigma]\nn\\
&=&-\frac{T}{f^{2}}V^{\mu}[V_{\mu}\partial_{\nu}\sigma-V_{\nu}\partial_{\mu}\sigma]\eea
Putting this together with (\ref{eom}), we find \be
-\frac{T}{f^{2}}V^{\mu}[V_{\mu}\partial_{\nu}\sigma-V_{\nu}\partial_{\mu}\sigma]=qU^{\mu}(F_{\nu\mu}+\frac{m}{q}S_{\nu\mu})\ee
or, \be
mV^{\mu}S_{\mu\nu}=qV^{\mu}F_{\nu\mu}+\frac{T}{f}V^{\mu}[V_{\mu}\partial_{\nu}\sigma-V_{\nu}\partial_{\mu}\sigma]\ee

Therefore, \be
V^{\mu}V_{\nu;\mu}-V^{\mu}V_{\mu;\nu}=\frac{q}{m}(V^{\mu}F_{\nu\mu}+\frac{mT}{qf}V^{\mu}[V_{\mu}\partial_{\nu}\sigma-V_{\nu}\partial_{\mu}\sigma])\ee
and, \be
{\dot{V}}_{\nu}=-f\partial_{\nu}f+\frac{q}{m}(V^{\mu}F_{\nu\mu}+\frac{mT}{qf}V^{\mu}[V_{\mu}\partial_{\nu}\sigma-V_{\nu}\partial_{\mu}\sigma])\ee
Defining a pure fluid factor, $N_{\mu\nu}$ as \be
N_{\mu\nu}=\frac{T}{f}(\sigma_{;\mu}V_{\nu}-\sigma_{;\nu}V_{\mu}),\ee
the acceleration vector simplifies to \be
{\dot{V}}_{\nu}=-f\partial_{\nu}f+\frac{q}{m}V^{\mu}(F_{\nu\mu}+\frac{m}{q}N_{\nu\mu}).\ee
Defining \be
G_{\mu\nu}=F_{\mu\nu}+\frac{m}{q}N_{\mu\nu},\label{gunify}\ee
substitution into (\ref{re1}) yields \bea
(\tilde{\Theta}+\frac{1}{f}V^{\mu}\partial_{\mu}f)_{;\gamma}V^{\gamma}&=&(-f\partial^{\alpha}f+\frac{q}{m}V_{\beta}G^{\alpha\beta})_{;\alpha}-\tilde{B}_{\alpha\beta}\tilde{B}^{\beta\alpha}\nn\\
&-&8 \pi t_{\mu \nu}V^{\mu}V^{\nu}-4\pi f^{2} t \eea which
simplifies to \bea
{\dot{\tilde{\Theta}}}+{\dot{\zeta}}&=&(-f\partial^{\alpha}f)_{;\alpha}+\frac{q}{m}V_{\beta}{}_{;\alpha}G^{\alpha\beta}+\frac{q}{m}V_{\beta}G^{\alpha\beta}{}_{;\alpha}\nn\\
&-& \tilde{B}_{\alpha\beta}\tilde{B}^{\beta\alpha}-8 \pi
t_{\mu\nu}V^{\mu}V^{\nu}-4\pi f^{2} t \eea where we have defined
$V^{\mu}\partial_{\mu}lnf =\zeta$. Using the relationship between
$B_{\mu\nu}$ and $\tilde{B}_{\mu\nu}$, we find \bea
{\dot{\tilde{\Theta}}}+{\dot{\zeta}}&=&(-f\partial^{\alpha}f)_{;\alpha}+\frac{q}{m}V_{\beta}{}_{;\alpha}G^{\alpha\beta}+\frac{q}{m}V_{\beta}G^{\alpha\beta}{}_{;\alpha}\nn\\
&-&
fB_{\alpha\beta}fB^{\beta\alpha}-2\tilde{B}^{\alpha\gamma}V_{\gamma}\partial_{\alpha}lnf
+\zeta^{2}-8 \pi t_{\mu\nu}V^{\mu}V^{\nu}-4\pi f^{2} t \nn\\
&=&\frac{q}{m}V_{\beta}{}_{;\alpha}G^{\alpha\beta}+\frac{q}{m}V_{\beta}G^{\alpha\beta}{}_{;\alpha}
-fB_{\alpha\beta}fB^{\beta\alpha}\nn\\
&+&\zeta^{2}-f^{2}(\partial^{2}lnf)-8 \pi
t_{\mu\nu}V^{\mu}V^{\nu}-4\pi f^{2} t \eea


In the limit $f\To1$ as $T\To0$ this reduces to the standard
Raychaudhuri equation . Simplifying the left hand side, \bea
{\dot{\tilde{\Theta}}}&=&\frac{q}{m}V_{\beta}{}_{;\alpha}G^{\alpha\beta}+\frac{q}{m}V_{\beta}G^{\alpha\beta}{}_{;\alpha}
-fB_{\alpha\beta}fB^{\beta\alpha}+\zeta^{2}\nn\\
&-&{\dot{\zeta}}-f^{2}(\partial^{2}lnf)-8 \pi
t_{\mu\nu}V^{\mu}V^{\nu}-4\pi f^{2} t \eea

Simplifying the $B_{\mu\nu}B^{\nu\mu}$ terms, we have \bea
{\dot{\tilde{\Theta}}}&=&\frac{q}{2m}S_{\alpha\beta}G^{\alpha\beta}+\frac{q}{m}V_{\beta}G^{\alpha\beta}{}_{;\alpha}
-(\tilde{\Sigma}_{\alpha\beta}\tilde{\Sigma}^{\beta\alpha}\nn\\
&+&\tilde{\Omega}_{\alpha\beta}\tilde{\Omega}^{\beta\alpha}+\frac{1}{3}\tilde{\Theta}^{2})+\zeta^{2}-{\dot{\zeta}}-f^{2}(\partial^{2}lnf)-8
\pi t_{\mu\nu}V^{\mu}V^{\nu}-4\pi f^{2} t \eea

This final form of the generalized Raychaudhuri equation shows all
the terms which reduce to the standard equation in the limit
$f(T)\longrightarrow1$.

\section {A Raychaudhuri equation for the unified non-Abelian
magneto fluid.} In a recent paper \cite{bambah}, we had generalized
the proposed unification of the Abelian magneto fluid \cite{mahajan}
to a non-Abelian magneto fluid. The equations of motion for the
magneto fluid clearly indicated the possibility of solitonic
solutions which are normally absent in the Abelian case. The
inherent non-linearities present in the Yang Mills magneto fluid
allow such possibilities and a natural question arises as to what will self gravity
do to the dynamics of a  non-Abelian magneto fluid. This is
particularly relevant in view of the compelling experimental
evidence from the relativistic heavy ion collider at Brookhaven
National Laboratory (BNL) that the universe in its first few moments
may have existed as a quark-gluon fluid. Since, large gravitational
fields are also present in this epoch, it provides us with a
motivation to give a generalization of the Raychaudhuri equation for
the non-Abelian magneto fluid.

The suggestion in \cite{bambah} was that each worldline would now
carry an internal index labelling the non-Abelian species.
Generalizing the Lorentz force law to the non-Abelian case, we had
derived the equation of motion for the magneto fluid in terms of a
unified antisymmetric tensor,
$M^{i}{}_{\mu\nu}=F^{i}{}_{\mu\nu}+\frac{m}{g}S^{i}{}_{\mu\nu}$,
where $F^{i}{}_{\mu\nu}$ is the standard Yang Mills field strength
tensor while $S^{i}{}_{\mu\nu}$ is given by:\be
S^{i}{}_{\mu\nu}=\partial_{\mu}(fU^{i}{}_{\nu})-\partial_\nu(fU^{i}{}_{\mu})-igf[A_{\mu},U_{\nu}]^{i}+igf[A_{\nu},U_{\mu}]^{i}-imf^{2}[U_{\mu},U_{\nu}]^{i}\ee
with the gauge covariant derivative being defined by $
{\cal{D}}_{\mu}=\partial_{\mu}-ig[A_{\mu}, ]$. The non-Abelian fluid equations of motion corresponding to Eqn.\ref{eom} for a "Yang-Mills Magneto-fluid", with entropy $\sigma$ are given by
\begin{equation}
T\p^{\nu}\sigma=gM^{\mu\nu}_{a}U_{a\mu}\label{naeom}.
\end{equation}

For a non-Abelian
magneto fluid velocity vector $U^{i}{}_{\mu}$, the deviation
equation can easily be written as \be
U^{i}{}^{\mu}{}_{;\nu\sigma}-U^{i}{}^{\mu}{}_{;\sigma\nu}=R^{\mu}{}_{\rho\nu\sigma}U^{i}{}^{\rho}.\ee
Examining $B^{i}{}_{\mu\nu}=U^{i}{}_{\mu;\nu}$, \be
B^{i}{}_{\mu\nu;\alpha}U_{i}{}^{\alpha}=(U^{i}{}_{\mu;\alpha}U_{i}{}^{\alpha})_{;\nu}-R_{\mu\rho\nu\alpha}U^{i}{}^{\mu}U_{i}{}^{\rho}-U^{i}{}_{\mu;\alpha}U_{i}{}^{\alpha}{}_{;\nu}\label{nadev}
.\ee From our earlier work \cite{bambah}, we write \be
tr{\dot{U}}_{\mu}=U^{i}{}_{\mu;\nu}U_{i}{}^{\nu}=\frac{g}{m}F^{i}{}_{\mu\nu}U_{i}{}^{\nu}\ee
for the generalization of the Lorentz force law in the particle
picture of the non-Abelian magneto fluid; the trace is over the
internal, gauge group indices ($i=1\dots N=$ dim(gauge group)).
Clearly, since now, $U^{i}{}_{\mu}U_{i}{}^{\mu}=-N$, we shall
henceforth, assume the $U^{i}{}_{\mu}$ are normalized so that
$U^{i}{}_{\mu}U_{i}{}^{\mu}=-1$, i.e; we assume
$U^{i}{}_{\mu}\longrightarrow U^{i}{}_{\mu}/\sqrt{N}$. With this
proviso, we can define an orthogonal projection tensor as before \be
h_{\mu\nu}=g_{\mu\nu}+U^{i}{}_{\mu}U_{i}{}_{\nu}.\ee It follows that
\be
U^{j}{}^{\mu}h_{\mu\nu}=U^{j}{}^{\mu}g_{\mu\nu}+U^{j}{}^{\mu}U^{i}{}_{\mu}U_{i}{}_{\nu}=U^{j}{}_{\nu}-\delta^{ij}U_{i}{}_{\nu}=0\ee
and \be
h_{\mu\nu}U^{j}{}^{\nu}=g_{\mu\nu}U^{j}{}^{\nu}+U^{i}{}_{\mu}U_{i}{}_{\nu}U^{j}{}^{\nu}=U^{j}{}_{\mu}-\delta^{ij}U_{i}{}_{\mu}=0\ee
where we have used \be U^{i}{}_{\mu}U^{j}{}^{\mu}=-\delta^{ij}\ee
from which it follows that $U^{i}{}_{\mu}U_{i}{}^{\mu}=-1$.
Equation(\ref{nadev}) can be reduced to \be
B^{i}{}_{\mu\nu;\alpha}U_{i}{}^{\alpha}=(\frac{g}{m}F^{i}{}_{\mu\alpha}U_{i}{}^{\alpha})_{;\nu}-R_{\mu\rho\nu\alpha}U^{i}{}^{\mu}U_{i}{}^{\rho}-U^{i}{}_{\mu;\alpha}U_{i}{}^{\alpha}{}_{;\nu}
.\ee Tracing over indices $\mu$ and $\nu$, and defining \be
\Theta^{i}=B^{i}{}_{\mu}{}^{\mu} ,\ee we find\be
{\dot{\Theta}}=(\frac{g}{m}F^{i}{}^{\mu}{}_{\alpha}U_{i}{}^{\alpha})_{;\mu}-R_{\rho\alpha}U^{i}{}^{\alpha}U_{i}{}^{\rho}-B^{i}{}_{\mu\alpha}B_{i}{}^{\alpha}{}^{\mu}\ee
Once again, we can decompose the tensor $B^{i}{}_{\mu\nu}$ into its
irreducible parts and write \be
B^{i}{}_{\mu\nu}=U^{i}{}_{\mu;\nu}=\Sigma^{i}{}_{\mu\nu}+\Omega^{i}{}_{\mu\nu}+\frac{1}{3}\Theta^{i}
h_{\mu\nu}.\ee It is to be noted here that we are considering a
special sector of the magneto fluid dynamics, that of a
non-interacting sector of the full non-Abelian fluid as is seen from
our definition of the acceleration vector
$a_{\mu}=U^{i}{}_{\mu;\nu}U_{i}{}^{\nu}$. From a fluid point of
view, we are dealing with a multi-species model of the non-Abelian
fluid. The full non-Abelian interactions of the fluid will be
explored in a future study. Returning to the decomposition of
$B^{i}{}_{\mu\nu}$, we write \bea
\Sigma^{i}{}_{\mu\nu}&=&\frac{1}{2}(U^{i}{}_{\mu;\nu}+U^{i}{}_{\nu;\mu})\nn-\frac{1}{3}\Theta^{i}h_{\mu\nu}\\
\Omega^{i}{}_{\mu\nu}&=&\frac{1}{2}(U^{i}{}_{\mu;\nu}-U^{i}{}_{\nu;\mu})\nn\\
\eea Clearly, these definitions give a decomposition of
$B^{i}{}_{\mu\nu}$ into its irreducible components. The resulting
Raychaudhuri equation, upon substitution of
$B^{i}{}_{\mu\nu}B_{i}{}^{\nu\mu}$ is, as expected, a multi-species
generalization of the standard Raychaudhuri equation.
 This  Non-Abelian generalization of Raychaudhuri's equation gives us a means of studying the shear and vorticity of quark-gluon
astrophysical plasmas.

\section{Conclusion}
High temperature plasmas have an important role to play in the early
universe. In what is known as the "classical" or the "radiation"
epoch of the universe, both gravitational and high temperature
effects are of equal importance. Thus for the evolutionary dynamics
of these hot gravitating plasmas, a generalization of the
Raychaudhuri equation to include finite temperature fluid forces as
well as electromagnetic effects was called for.  Using a unified
magneto-fluid approach to construct a generalized Raychaudhuri
equation,  we have attempted to respond to this call. Non-Linear
plasmas also find an application in cosmology through the (now
compelling) evidence that a non Abelian, non linear quark gluon
fluid existed in the early epochs of the universe. To deal with the
evolution of this fluid when the gravitational effects are strong ,
we have laid the foundation of a generalized Raychaudhuri equation
for the evolution of non Abelian gravitating plasmas. Most
interesting  plasmas involve collective effects which are non-linear
even in the special relativistic situation and the inclusion of
gravity can only lead to more intriguing highly non linear
phenomena. The investigation of further implications of the ideas
germinated in this paper is a promising avenue for future research.
Amalgamation and unification of ideas that cut across barriers
always enriches physics and we feel that any work of this kind is a
fitting tribute to a man whose physics cut across international
boundaries.

\end{document}